%% file: articolo5.tex
\documentclass[twocolumn,aps,prl,showpacs,groupedaddress]{revtex4-1}

\usepackage{epsfig,amssymb,amsmath,mathrsfs,color}
\usepackage[hypertex,linkcolor=red]{hyperref}
\graphicspath{{imma/}}
\input{myQcircuit}

\newcommand{\Det}{\operatorname{Det}}
\newcommand{\su}{\mathbf{su}}
\newcommand{\SU}{\mathbf{SU}}
\newcommand{\so}{\mathbf{so}}
\newcommand{\SO}{\mathbf{SO}}
\renewcommand{\O}{\mathbf{O}}

\newtheorem{lemma}{Lemma} 
 \newtheorem{theorem}{Theorem}
 
 \def\Proof{\medskip\par\noindent{\bf Proof.
  }}
\def\qed{$\,\blacksquare$\par} 
 \def\>{\rangle}
\def\<{\langle}

\begin{document}

\title{Universality of Computation in Real Quantum Theory}
\author{Alessio Belenchia}\email{alessio.belenchia@gmail.com}
\affiliation{SISSA, Via Bonomea 265, 34136, Trieste, Italy}
\author{Giacomo Mauro D'Ariano}\email{dariano@unipv.it}
\affiliation{QUIT Group, Dipartimento di Fisica, Universit\`a di Pavia, via Bassi 6, 27100 Pavia, Italy} 
\homepage{http://www.qubit.it}
\affiliation{Istituto Nazionale di Fisica Nucleare, Gruppo IV, via Bassi 6, 27100 Pavia, Italy} 
\author{Paolo Perinotti}\email{paolo.perinotti@unipv.it} 
\affiliation{QUIT Group, Dipartimento di Fisica, Universit\`a di Pavia, via Bassi 6, 27100 Pavia, Italy} 
\homepage{http://www.qubit.it}
\affiliation{Istituto Nazionale di Fisica Nucleare, Gruppo IV, via Bassi 6, 27100 Pavia, Italy} 
\date{\today}
\begin{abstract}
  Recently de La Torre et al.\cite{mms} reconstructed Quantum Theory from its local structure on the
  basis of local discriminability and the existence of a one-parameter group of bipartite
  transformations containing an entangling gate. This result relies on universality of an entangling
  gate for quantum computation. Here we prove universality of C-NOT with local gates for Real
  Quantum Theory (RQT), showing that such universality would not be sufficient for the result,
  whereas local discriminability and the qubit structure play a crucial role.  For reversible
  computation, generally an extra rebit is needed for RQT. As a byproduct we also provide a short
  proof of universality of C-NOT for CQT.
\end{abstract}
\maketitle
In recent years Quantum Information has spawned an unprecedented revival of interest in quantum
foundations, providing original lines of search based on the surprising power of Quantum Theory as a
model for information processing. This has led many authors to believe that ``information'' is the
key to the solution of the mystery of quantum mechanics \cite{brassard,fuchs}. Along these lines the
seminal work of Hardy \cite{hardy1} has opened the route to the new axiomatization program
\cite{bruckner,masanes,hardy2,darCUP}, including the derivation of the theory from
information-theoretical principles \cite{infoder,viewpo}.

Some of the attempts at an informational axiomatization explored the possibility of deriving the
bipartite correlations of the theory from the local qubit structure \cite{Barnum}, however with the
inclusions of spurious correlations for more than two systems. Ref. \cite{mms} has then
reconstructed quantum theory in this way, with the addition of local discriminability and the
existence of a one-parameter group of bipartite transformations containing an entangling gate.  For
the derivation of this result the universality of entangling gates for quantum computation
\cite{harrow, bry} plays a crucial role. 

The existence of a universal gate set with a single bipartite gate is characteristic of quantum
computation, as opposed to the classical one \cite{deu1,barenco,divincenzo}. Since universality of a
bipartite gate plays a crucial role in the result of Ref. \cite{mms}, one may wonder if it is
specific only of Quantum Theory, or it holds instead also for other probabilistic theories, in the
absence of the requirements of local discriminability and the local qubit structure, as is the case
e.g.~of RQT.  Local discriminability, in particular, is an essential feature of a probabilistic
theory for multipartite systems (for a thorough exploration of local tomography, which is an
equivalent formulation of local discriminability, see Ref. \cite{woot}).

In the present letter we will prove that universality of C-NOT with local gates holds indeed also
for RQT. Differently from Complex Quantum Theory (CQT), for RQT generally an extra rebit is needed
for reversible computation.  We formulate universal computation with a single bipartite gate as an
informational axiom in the context of general probabilistic theories, then focusing on CQT and RQT
only, and providing simple proofs of universality for both theories. The simplified proof is useful
also in the complex case, since it provides a much shorter derivation than the original ones
\cite{deu1,barenco,divincenzo,bry}. In the real case, an interesting feature pops up, which is the
requirement of a single overhead rebit for the circuit implementation of arbitrary orthogonal (i.e.
real unitary) transformations.




\medskip
We say that a general probabilistic theory admits computation with a {\em strongly universal}
bipartite gate if every reversible transformation of $N$ elementary systems (i.e. bits, qubits,
rebits, etc.) can be perfectly simulated by a circuit of $N$ elementary systems made only of local
reversible transformations and sufficiently many uses of the bipartite gate.  We say that the theory
admits a {\em weakly universal} bipartite gate if every reversible transformation of $N$ elementary
systems can be perfectly simulated by a circuit of $N+p(N)$ elementary systems made only of local
reversible transformations and sufficiently many uses of a single bipartite gate, discarding the
auxiliary $p(N)$ systems, where $p(x)$ is a polynomial in $x$.

We provide now a simplified proof that the C-NOT is strongly universal for computation in CQT.

\medskip
The elementary system in quantum computation is the qubit. The Hilbert space for a
register of $N$ qubits is $\mathbb{C}^{2^{N}}$, and its reversible transformations form
the Lie group is $\SU(2^{N})$.
Every element of $\SU(2^N)$ is the exponential of an anti-Hermitian operator. For a single qubit the
group $\SU(2)$ has the following generators
\begin{align}
  X:=
  \begin{pmatrix}
    0 & 1  \\
    1 & 0
  \end{pmatrix},\ 
  Z:=
  \begin{pmatrix}
    1 & 0 \\
    0 & -1
  \end{pmatrix},\ 
  Y:=
  \begin{pmatrix}
    0 & -i  \\
    i & 0
  \end{pmatrix}
\end{align}
We also introduce the bipartite C-NOT gate $V=V^\dag$ in $\SU(4)$
\begin{align}
  &V|i\>|j\>:=|i\>|i\oplus j\>,
\end{align}
where $|i\>$ is an element of the computational basis $\{|0\>,|1\>\}\subset\mathbb C^2$, while
$\oplus$ denotes the sum modulo 2. The qubit on the left is named {\em control} and the qubit on the
right is named {\em target}. Speaking about universality, one may think that the gate $\bar V$ with
the target and the control exchanged, is different from the gate $V$, however, in this spirit, one
can also notice that $\bar V$ is obtained from $V$ using local gates as follows
\begin{equation}
\bar V:=(H\otimes H)V(H\otimes H),
\end{equation}
where $H$ is the Hadamard gate
\begin{equation}
H=\frac{1}{\sqrt2} \begin{pmatrix} 1 &1\\ 1 &-1
\end{pmatrix}.
\end{equation}
The bipartite swap gate $P|\phi\>|\psi\>=|\psi\>|\phi\>$ can be
obtained from the C-NOT gate $V$ as $P=V\bar V V$

When multiple qubits are involved in the computation, we will denote by $V_{ij}$ the C-NOT where the
$i$th qubit is the control and $j$th qubit is the target. 

In the following we will denote by ${\mathcal L}$ the following basis for the
Lie algebra $\su(2^N)$ of the group $\SU(2^N)$
\begin{equation}
{\mathcal L}_N=\{L_1\otimes L_2\otimes\ldots\otimes L_N\}\backslash\{I^{\otimes N}\},\;
L_j\in\{I,X,Y,Z\}.
\end{equation}
The special case in which only one $L_j$ for fixed $j$ is different from the identity corresponds to
the basis for the Lie algebra of the local gates of the $j$th qubit.

We now prove some preliminary lemmas which are needed by the main theorem.
\begin{lemma}\label{l1} Starting from the element $I^{\otimes(N-1)}\otimes X$ one can generate the
  whole basis $\mathcal L_N$ using only C-NOTs and local gates.
\end{lemma}
\Proof Using the following trivial identity
\begin{equation}
V( I\otimes X)V^\dag=X \otimes X,\quad P( I\otimes X)P^\dag=X \otimes I,
\end{equation}
we can generate all strings in ${\mathcal L}_N$ with $L_j\in\{I,X\}$ by applying $V_{jN}$. With local
gates we can then generate the whole ${\mathcal L}_N$.\qed

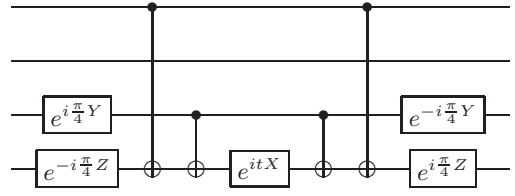
\begin{figure}
\begin{equation*}
    \Qcircuit @C=1em @R=.7em @! R { & \qw&\ctrl{3}&\qw&\qw&\qw&\ctrl{3}&\qw&\qw\\
      &\qw&\qw&\qw&\qw&\qw&\qw&\qw&\qw\\
      &\gate{e^{i\frac\pi4 Y}}&\qw&\ctrl{1}&\qw&\ctrl{1}&\qw&\gate{e^{-i\frac\pi4 Y}}&\qw\\
      &\gate{e^{-i\frac\pi4 Z}}& \targ&\targ&\gate{e^{itX}}&\targ&\targ&\gate{e^{i\frac\pi4 Z}}&\qw}
\end{equation*}
\caption{Relization of the multipartite unitary transformation $U=\exp(it X\otimes I\otimes Z\otimes Y)$
  using only C-NOTs and local gates corresponding to rotations of $\pm\pi/2$ around the $Y$ and $Z$ axes.}\label{figU}
\end{figure}

As an example of realization of gate according to Lemma \ref{l1} is given in Fig. \ref{figU}.

We thus proved that with local gates and the two-bit entangling gate C-NOT we can obtain all gates
of the form $\exp(it\Lambda)$, with $\Lambda\in{\mathcal L}_N$. By repeated applications of such gates
for varying $t$ and $\Lambda$ we generate the subgroup ${\mathbf H}\subseteq\SU(2^N)$. 

We now have the following lemma.
\begin{lemma} The subgroup $\mathbf H\subseteq\SU(2^N)$ is dense in $\SU(2^N)$.
\end{lemma}
\Proof The statement is an immediate consequence of the Lie-Trotter formula
\begin{equation}
  e^{a\Lambda_{1}+b\Lambda{2}}=\lim_{n\rightarrow\infty}\left(e^{\frac{a\Lambda_{1}}{n}}e^{\frac{b\Lambda_{2}}{n}}\right)^{n},
\end{equation}
where convergence is to be considered in the strong topology \cite{ichinose}.\qed
The last lemma that we need is the following
\begin{lemma}[Brylinski \cite{bry}]\label{dense}
  Let \textbf{G} be a compact Lie group. If $H_{1}$,$\cdots$,$H_{k}$ are closed connected subgroups
  and they generate a dense subgroup of \textbf{G}, then in fact they generate \textbf{G}.
\end{lemma}
We now have all elements for proving our first main theorem
 \begin{theorem}[Strong universality of C-NOT] The C-NOT gate is
  strongly universal for quantum computation.
\end{theorem}
\Proof We observe that for each $\Lambda\in{\mathcal L}$ the one parameter subgroup of $\SU(2^N)$
$\left\{e^{i\Lambda t},\ t\in[0,2\pi)\right\}$ is closed and connected. Then we apply Lemma
\ref{dense} where the groups $H_k$ are the one-parameter Lie groups obtained by exponentiating
each element of ${\mathcal L}_N$. \qed

\medskip
We now prove universality for RQT. This theory shares a lot of features with CQT, and in some sense
is a subset of it. Nevertheless it has also some important differences from CQT, mostly that it
doesn't posses local discriminability, but has only the bilocal one \cite{woot,purif}.

The group of reversible transformations on $\mathbb{R}^{2^N}$, i.e.
transformations that preserve the norm of vectors, is the ortogonal
group $\O(2^{N})$ that is a compact not connected Lie group.  Now we
want to prove that the C-NOT (which is an ortogonal operator) and
local gates are sufficient to generate all the gates in $\SO(2^{N})$.

Notice that the $\bar V$ gate can still be obtained from
the C-NOT with local $\SO(2)$ gates as follows
\begin{equation}
\bar V=(\tilde Y\tilde H\otimes\tilde H)V(\tilde H\tilde Y\otimes\tilde H)
\end{equation}
where
\begin{equation}  \tilde Y:=  \begin{pmatrix}    0 & 1  \\    -1 & 0  \end{pmatrix},\;
  \tilde{H}:=\frac{1}{\sqrt{2}}
  \begin{pmatrix}
    1 & -1 \\
    1 & 1
  \end{pmatrix}\in \SO(2).
\end{equation}
Hence we also get the SWAP gate $P=V\bar V V$. We now prove universality along lines analogous to
the proof for CQT. We will need to consider the transformations of $\O(2^{N})$ with determinant equal
to $-1$ separately, because these cannot be obtained via the exponential map as before.

\medskip Let us start with the first task, i.e. obtain all $\SO(2^{N})$ from C-NOT and local gates.
Since every orthogonal matrix is the exponential of an antisymmetric matrix, the generators of
$\so(2^{N})$ are strings of $\tilde Y,X,Z,I$, with the constraint that they are antisymmetric. It is
easy to verify that this amounts to require that they contain an odd number of $\tilde Y$.

For the case of two rebits the generators of local gates are then
$I\otimes \tilde Y$ and $\tilde Y\otimes I$.  If we act on these
generators with C-NOT and SWAP we obtain
\begin{equation}
  Z\otimes \tilde Y,\quad\tilde Y\otimes X,\quad X\otimes \tilde Y,\quad \tilde Y\otimes Z,
\end{equation}
namely we have the six generators of the $\so(4)$ algebra. The
induction hypothesis is now that we can construct an arbitrary string
of length $N-1$ with the constraint of containing only an odd number
of $\tilde Y$'s, and we have to prove that we can construct an
arbitrary string of length $N$ using C-NOT. By hypotesis we have the
following generators
\begin{equation}\nonumber
  I\otimes \tilde Y\otimes B,\quad I\otimes X\otimes A,\quad I\otimes Z\otimes A,
\end{equation}   
where $A$ is an arbitary string of length $N-2$ with an odd number of
$\tilde Y$ and $B$ is an arbitary string of length $N-2$ with an even
number of $\tilde Y$. Acting on these operators with C-NOT and SWAP we
obtain
\begin{equation}\nonumber
  Z\otimes \tilde Y\otimes B,\quad X\otimes X\otimes A.
\end{equation}
Now we can replace $Z$ with $X$ and viceversa by acting with the local
gate $\tilde{H}$ modulo a sign on $Z$ (the sign is not relevant, since we are considering
Lie-algebra elements).

Finally, acting with C-NOT on $X\otimes Z\otimes A$ we obtain $\tilde
Y\otimes \tilde Y\otimes A$. This conclude the induction proof, and we
can see that the whole group $\SO(2^{N})$ is generated using Lemma
\ref{dense} in the same way as before.


But in this case we still have a problem. The Lie algebra $\so(2^{N})$
can generate by exponentiation the connected component of the
orthogonal group, i.e. the special ortogonal group, but it is
impossible to construct from local gates and C-NOT a gate that has
determinant equal to $-1$. This is because if we consider a local gate
with determinant $-1$ or also the C-NOT gate and take the tensor
product with the identity or another unit determinant gate we always
obtain a gate with determinant $+1$. This follows directly from the
property of the Kronecker product, i.e if $A\in \O(2^{N})$ and $B\in
\O(2^M)$ then 
\begin{equation}\label{dettensor}
\Det(A\otimes B)=\Det(A)^{2^M}\Det(B)^{2^N}.
\end{equation}

But if one wants to construct an $N$-rebits gate $S$ with determinant $-1$, he can instead use an
ancillary rebit and construct the $N+1$-rebits gate $I\otimes S$ using C-NOT and local gates (notice
that by Eq. (\ref{dettensor}) $\Det(S)=-1$, whereas $\Det(I\otimes S)=1$). We
have thus proved the weak-universality of local gates and C-NOT for RQT. 

\medskip In this work we have seen that in RQT local gates and C-NOT,
are universal for reversible computation, as in CQT, but an additional
ancillary rebit is needed for universality of RQT. Using a similar
line of proof we have also provided a very simple and short proof of
universality for CQT. It is argued that RQT has a weak-universality
property due to the fact that it does not satisfy local
discriminability, and it is interesting to ask if the universality
property is a good axiom for CQT in the presence causality and local
discriminability.

\end{document}

%% file: myQcircuit.tex
\usepackage[matrix,frame,arrow]{xy}
\usepackage{amsmath}

\newcommand{\qw}[1][-1]{\ar @{-} [0,#1]}
\newcommand{\qwx}[1][-1]{\ar @{-} [#1,0]}

\newcommand{\gate}[1]{*{\xy *+<.6em>{#1};p\save+LU;+RU **\dir{-}\restore\save+RU;+RD **\dir{-}\restore\save+RD;+LD **\dir{-}\restore\POS+LD;+LU **\dir{-}\endxy} \qw}

\newcommand{\control}{*!<0em,.025em>-=-{\bullet}}

\newcommand{\ctrl}[1]{\control \qwx[#1] \qw}

\newcommand{\targ}{*!<0em,.019em>=<.79em,.68em>{\xy {<0em,0em>*{} \ar @{ - } +<.4em,0em> \ar @{ - } -<.4em,0em> \ar @{ - } +<0em,.36em> \ar @{ - } -<0em,.36em>},<0em,-.019em>*+<.8em>\frm{o}\endxy} \qw}
\newcommand{\Qcircuit}[1][0em]{\xymatrix @*=<#1>} 

